%
%
%
\def\today{\ifcase\month\or January\or February\or March\or April\or May\or
June\or July\or August\or September\or October\or November\or December\fi
\space\number\day, \number\year}
%
%
\newcount\notenumber

\def\note{\global\advance\notenumber by 1 \footnote{$^{\the\notenumber}$}}
%
%
\newif\ifsectionnumbering
\newcount\eqnumber
\def\cleareqnumber{\eqnumber=0}
\def\numbereq{\global\advance\eqnumber by 1
\ifsectionnumbering\eqno(\the\secnumber.\the\eqnumber)\else\eqno
(\the\eqnumber)\fi}
\def\eqalinno{{\global\advance\eqnumber by 1}
\ifsectionnumbering(\the\secnumber.\the\eqnumber)\else(\the\eqnumber)\fi}
\def\name#1{\ifsectionnumbering\xdef#1{\the\secnumber.\the\eqnumber}
\else\xdef#1{\the\eqnumber}\fi}
\def\nosectionnumbering{\sectionnumberingfalse}
\sectionnumberingtrue
%
%
\newcount\refnumber

\immediate\openout1=refs.tex
\immediate\write1{\noexpand\frenchspacing}
\immediate\write1{\parskip=0pt}
\def\ref#1#2{\global\advance\refnumber by 1%
[\the\refnumber]\xdef#1{\the\refnumber}%
\immediate\write1{\noexpand\item{[#1]}#2}}
\def\tie{\noexpand~}

%
%
\font\twelvebf=cmbx10 scaled \magstep1
\newcount\secnumber

\def\newsection#1.{\ifsectionnumbering\cleareqnumber\else\fi%
	\global\advance\secnumber by 1%
	\bigbreak\bigskip\par%
	\line{\twelvebf \the\secnumber. #1.\hfil}\nobreak\medskip\par\noindent}
%
%
%
\def \sqr#1#2{{\vcenter{\vbox{\hrule height.#2pt
	\hbox{\vrule width.#2pt height#1pt \kern#1pt
		\vrule width.#2pt}
		\hrule height.#2pt}}}}

%
%
%
\newdimen\fullhsize
\def\fiddle{\fullhsize=6.5truein \hsize=3.2truein}
\def\fullline{\hbox to\fullhsize}
\def\mkhdline{\vbox to 0pt{\vskip-22.5pt
	\fullline{\vbox to8.5pt{}\the\headline}\vss}\nointerlineskip}
\def\mkftline{\baselineskip=24pt\fullline{\the\footline}}
\let\lr=L \newbox\leftcolumn
\def\twocolumns{\fiddle
	\output={\if L\lr \global\setbox\leftcolumn=\columnbox
		\global\let\lr=R \else \doubleformat \global\let\lr=L\fi
		\ifnum\outputpenalty>-20000 \else\dosupereject\fi}}
\def\doubleformat{\shipout\vbox{\mkhdline
		\fullline{\box\leftcolumn\hfil\columnbox}
		\mkftline} \advancepageno}
\def\columnbox{\leftline{\pagebody}}
\nosectionnumbering
\magnification=1200
\def\pr#1 {Phys. Rev. {\bf D#1\tie }}
\def\pe#1 {Phys. Rev. {\bf #1\tie}}
\def\pre#1 {Phys. Rep. {\bf #1\tie}}
\def\pl#1 {Phys. Lett. {\bf #1B\tie }}
\def\prl#1 {Phys. Rev. Lett. {\bf #1\tie }}
\def\np#1 {Nucl. Phys. {\bf B#1\tie }}
\def\ap#1 {Ann. Phys. (NY) {\bf #1\tie }}
\def\cmp#1 {Commun. Math. Phys. {\bf #1\tie }}
\def\imp#1 {Int. Jour. Mod. Phys. {\bf A#1\tie }}
\def\mpl#1 {Mod. Phys. Lett. {\bf A#1\tie}}
\def\cqg#1 {Class. Quantum Grav. {\bf #1\tie }}
\def\tie{\noexpand~}

\def\d{\delta(\sigma-\sigma ')}
\def\dpr{\delta'(\sigma-\sigma ')}

\parskip=15pt plus 4pt minus 3pt
\headline{\ifnum \pageno>1\it\hfil Anisotropic Null String
Cosmologies $\ldots$\else \hfil\fi}
\font\title=cmbx10 scaled\magstep1
\font\tit=cmti10 scaled\magstep1
\footline{\ifnum \pageno>1 \hfil \folio \hfil \else
\hfil\fi}
\raggedbottom


\overfullrule0pt


\rightline{\vbox{\hbox{NYU-TH/98/10/01}\hbox{RU98-11-B}\hbox{gr-qc/9810062}}}
\vfill
\centerline{\title ANISOTROPIC NULL STRING COSMOLOGIES}
\vfill
{\centerline{\title Ioannis Giannakis${}^{a,b}$,
K. Kleidis${}^{c}$, A. Kuiroukidis${}^{c}$ and
D. Papadopoulos${}^{c}$ \footnote{$^{\dag}$}
{\rm e-mail: \vtop{\baselineskip12pt
\hbox{giannak@theory.rockefeller.edu, kleidis@astro.auth.gr,}
\hbox{kuirouki@astro.auth.gr, papadop@astro.auth.gr}}}}
}
\medskip
\centerline{$^{(a)}${\tit Department of Physics, New York University}}
\centerline{\tit 4 Washington Pl., New York, NY 10003}
\medskip
\centerline{$^{(b)}${\tit Physics Department, The Rockefeller
University}}
\centerline{\tit 1230 York Avenue, New York, NY
10021-6399}
\medskip
\centerline{$^{(c)}${\tit Department of Physics}}
\centerline{\tit Section of Astrophysics, Astronomy and Mechanics}
\centerline{\tit Aristotle University of Thessaloniki,
54006 Thessaloniki, Greece} 
\vfill
\centerline{\title Abstract}
\bigskip
{\narrower\narrower
We study string propagation in an anisotropic, cosmological
background. We solve the equations of motion and the constraints
by performing a perturbative expansion of the string coordinates
in powers of $c^2$, the world-sheet speed of light. To zeroth order
the string is approximated by a tensionless string (since $c$ is
proportional to the string tension $T$). We obtain exact, analytical
expressions for the zeroth and the first order solutions and we discuss
some cosmological implications.
\par}
\vfill\vfill\break


\newsection Introduction.%
Strings in curved spacetimes provide the best framework for
studying gravity in the context of string theory.
The action for a bosonic string propagating in a $D$-dimensional
Riemannian manifold is given by
$$
S=-{T\over 2}\int d\tau d\sigma \sqrt{-h}h^{\alpha \beta }
(\tau ,\sigma )
G_{\mu\nu}(X)\partial _{\alpha}X^{\mu} \partial _{\beta}X^{\nu}
\numbereq\name{\eqascuit}
$$
where $\mu, \nu = 0,1,...(D-1)$ are spacetime indices,
$\alpha ,\beta = 0,1$
are worldsheet indices and $T = (2\pi \alpha ^{\prime })^{-1}$ is the
string tension. The metric $h^{\alpha \beta }
(\tau ,\sigma )$ describes the intrinsic geometry of the two-dimensional
worldsheet while $G_{\mu\nu}(X)$ describes the geometry of the
$D$-dimensional target space.
This particular action does not permit us to discuss
the limit $T \mapsto 0$. Following the analogous
massless particle case \ref{\clsa}{A. Karlhede and U. Lindstrom,
\cqg3 (1986) L3} we
are led to a reformulated action which in the
conformal gauge $ (h_{\alpha \beta }(\tau ,\sigma ) 
=exp[\phi (\tau ,\sigma )]\eta _{\alpha \beta })$,
where $\phi (\tau ,\sigma)$ is an arbitrary function and
$\eta _{\alpha \beta }=diag(-1,+1)$, leads to the
classical equations of motion
$$
\ddot{X}^{\mu}-c^{2}(T)(X^{\mu})^{\prime \prime}
+\Gamma ^{\mu}_{\nu\rho}(X)
[\dot{X}^{\nu}\dot{X}^{\rho}
-c^{2}(T)X^{\prime{\nu}}X^{\prime{\rho}}] = 0
\numbereq\name{\eqanvh}
$$
with $c=2{\lambda}T$ ($\lambda$ is a Lagrange multiplier)
and $\Gamma ^{\mu}_{\nu\rho}(X)$ are the Christofell symbols associated
to the metric $G_{\mu\nu}(X)$. Variation of this action with respect to the
two-dimensional metric $h^{\alpha \beta }(\tau ,\sigma )$ also yields
two constraints
$$
G_{\mu\nu}(X)[\dot{X}^{\mu}\dot{X}^{\nu}+c^{2}(T)
X^{\prime{\mu}}X^{\prime{\nu}}] = 0,
\quad G_{\mu\nu}(X)\dot{X}^{\mu}X^{\prime{\nu}} = 0.
\numbereq\name{\eqaxsiuof}
$$
Dot and prime stand for differentiation with respect to $\tau$
and $\sigma$ respectively.
Different perturbative approaches were
introduced in order to solve the EOM and
the constraints of strings propagating in curved backgrounds:
the center of mass expansion \ref{\veg}{H. J. de Vega and N. Sanchez,
\pl197 (1987) 320; \np309 (1988) 552; \np309 (1998) 557;
\pr42 (1990) 3269.}, the $\tau$-expansion \ref{\ven}{M. Gasperini,
N. Sanchez and G. Veneziano, \imp6 (1991) 365.}, the null string
expansion \ref{\nik}{H. J. de Vega and A. Nicolaidis, \pl295 (1992)
214, H. J. de Vega, A. Nicolaidis and I. Giannakis, \mpl10 (1995) 2479,
C. Lousto and N. Sanchez, \pr54 (1996) 6399, M. Dabrowski and
A. Zheltukhin, {\it Perturbative String Dynamics Near the Photon Sphere},
hep-th/9809176.}
while in \ref{\nva}{F. Combes, H. J. de Vega,
A. V. Mikhailov and N. Sanchez, \pr50 (1994) 2754, H. J. de Vega,
A. L. Larsen and N. Sanchez, \np427 (1994) 643, H. J. de Vega and
L. I. Egusquiza, \pr49 (1994) 763, A. L. Larsen and N. Sanchez,
\pr50 (1994) 7493.} inverse scattering methods and soliton
techniques were introduced in order to obtain exact solutions.
Solutions in closed form have also been found for some
relevant gravitational backgrounds in \ref{\nvbo}
{H. J. de Vega and N. Sanchez,
\np317 (1989) 3269, D. Amati and C. Klimcik, \pl210 (1992) 92.}.
The null string method consists of inserting
into Eqns. (\eqanvh) and (\eqaxsiuof) the expansion
$$
X^{\mu }(\tau ,\sigma )=A^{\mu }(\tau ,\sigma )+c^{2}B^{\mu }(\tau ,\sigma )
+O(c^{4}), \qquad  \mid B^{\mu }\mid \ll \mid A^{\mu }\mid.
\numbereq\name{\eqaohg}
$$
To zeroth order in $c^2$ we find
$$
\ddot{A}^{\rho }+\Gamma ^{\rho }_{\kappa \lambda }
\dot{A}^{\kappa }\dot{A}^{\lambda }=0, \quad
\dot{A}^{\mu }\dot{A}^{\nu }G_{\mu \nu }=0, \quad
\dot{A}^{\mu }A^{'\nu }G_{\mu \nu }=0.
\numbereq\name{\eqavnp}
$$
We observe that the coordinates $A^{\mu}(\tau, \sigma)$
describe a null string
\ref{\schild}{A. Schild, \pr16 (1977) 1722.}, a collection
of points moving independently along null geodesics. The
second constraint ensures that each point of the string
propagates in a direction perpendicular to the string.
The next order correction $B^{\mu}(\tau, \sigma)$
(first order in $c^2$) obeys the EOM
$$
\ddot{B}^{\mu }+\Gamma ^{\mu }_{\alpha \beta }
(\dot{A}^{\alpha }\dot{B}^{\beta }
+\dot{B}^{\alpha }\dot{A}^{\beta })+\Gamma ^{\mu }_{\alpha \beta ,\nu }
\dot{A}^{\alpha }\dot{A}^{\beta }B^{\nu }=A^{''\mu }
+\Gamma ^{\mu }_{\kappa \lambda}A^{'\kappa }A^{'\lambda }
\numbereq\name{\eqrug}
$$
supplemented with the constraints
$$
\eqalign{
2\dot{A}^{\mu }\dot{B}^{\nu }G_{\mu \nu }
+\dot{A}^{\mu }\dot{A}^{\nu }B^{\rho }G_{\mu \nu ,\rho }
+A^{'\mu }A^{'\nu }G_{\mu \nu}&=0\cr
(\dot{A}^{\mu }B^{'\nu }+\dot{B}^{\mu }A^{'\nu })G_{\mu \nu }+
\dot{A}^{\mu }A^{'\nu }B^{\rho }G_{\mu \nu ,\rho }&=0 \cr}
\numbereq\name{\eqroi}
$$
where $G_{\mu \nu ,\rho }$ and $\Gamma ^{\mu }_{\alpha \beta ,\rho }$
indicate differentiation with respect to $A^{\rho}$.
We note that in this approximation scheme we expand around
the null string, an extended object as opposed to the
center of mass expansion in which the zeroth order solution
corresponds to the center of mass of the string, a pointlike structure.

\newsection Null Strings in Milne Spacetimes.
In this section we will apply the null string approach to
solve the EOM and the constraints of a string propagating
in a Milne
cosmological background. Null strings propagating in various
spacetime geometries including anisotropic models, were previously
considered in \ref{\nvco}{S. Kar, \pr53 (1996) 6842, M. Dabrowski
and A. Larsen, \pr55 (1997) 6409, P. Porfyriadis and D. Papadopoulos,
\pl417 (1998) 27, M. Dabrowski and A. Larsen, \pr57 (1998) 5108,
S. Roshchupkin and A. Zheltukhin, \cqg12 (1995) 2519.}.
The line element of the Milne spacetime is defined as
$$
ds^{2}=-(dX^{0})^{2}+(dX^{1})^{2}+(dX^{2})^{2}+(X^{0})^{2}(dX^{3})^{2}.
\numbereq\name{\eqziuc}
$$
Although Milne spacetimes are merely unconventional coordinatizations
of flat geometries analogous to the Rindler metrics, they lead to
non-trivial effects.
The equations of motion and the constraints
for $A^{\mu}$, Eqn. (\eqavnp) become
$$
\eqalign{
&\ddot{A}^{0}+(A^{0})(\dot{A}^{3})^{2}=0, \quad \ddot{A}^{1}
=\ddot{A}^{2}=0, \quad
\ddot{A}^{3}+{{2{\dot{A}^{0}}{\dot{A}^{3}}}\over {A^{0}}}=0\cr
&(\dot{A}^{0})^{2}=(\dot{A}^{1})^{2}+(\dot{A}^{2})^{2}
+(A^{0})^{2}(\dot{A}^{3})^{2}, \quad
(\dot{A}^{0}A^{'0})=(\dot{A}^{1}A^{'1})
+(\dot{A}^{2}A^{'2})+(A^{0})^{2}
(\dot{A}^{3}A^{'3}).\cr}
\numbereq\name{\eqbmoc}
$$
There are two classes of solutions:
$$
\eqalign{
&(A^{0})^{2}(\tau, \sigma)=((P^{1})^{2}+(P^{2})^{2})(\sigma)\tau ^{2}
+2P^{0}(\sigma)\tau +I^{0}(\sigma), \quad
A^{1,2}(\tau, \sigma)=P^{1,2}(\sigma)\tau +I^{1,2}(\sigma)\cr
&A^{3}(\tau, \sigma)={1\over 2}ln{{\left({(P^{0}+P^{3})\tau+I^{0}}\over
{(P^{0}-P^{3})\tau+I^{0}}\right)}} +I^{3}(\sigma)\cr}
\numbereq\name{\eqohdp}
$$
provided that $\Delta=I^{0}((P^{1})^{2}+(P^{2})^{2})-{P^{0}}^2 < 0$
and $P^{\mu}, I^{\mu}$ are arbitrary functions of $\sigma$ that
obey the constraints
$$
(P^{0})^{2}=I^{0}((P^{1})^{2}+(P^{2})^{2})+(P^{3})^{2}, \quad
P^{1}(I^{1})^{'}+P^{2}(I^{2})^{'}+P^{3}(I^{3})^{'}
={{P^0}{I^0}'\over {2I^0}}.
\numbereq\name{\eqxbvo}
$$
These constraints tie together the initial momentum $P^{\mu}(\sigma)$
and shape of the string $I^{\mu}(\sigma)$.
When the worldsheet timelike variable vanishes
$(\tau =0)$, then the cosmic time variable
$(A^{0})$ assumes finite value ${\sqrt {I^{0}}}$
and the position of the string is specified by $(I^{1}(\sigma), I^{2}(\sigma),
I^{3}(\sigma))$. As
the string evolves $(\tau \rightarrow +\infty )$, and
$(A^{0}\rightarrow +\infty )$, we notice that $A^{3}\rightarrow A^{3}_{max}$
$$
A^{3}_{max}={1\over 2}ln\left({{P^{0}+P^{3}}\over {P^{0}-P^{3}}}\right)=
ln\left(\rho +\sqrt{1+\rho ^{2}}\right)
\numbereq\name{\eqasnco}
$$
with $\rho={{P^{3}}\over {{\sqrt{((P^1)^2+(P^2)^2)I^{0}}}}}$.
The string bounces at a finite distance from the origin as it moves towards 
the direction of the anisotropy. One needs increasingly higher values
of the momentum $P^{3}$
in order to penetrate into higher values of the direction of anisotropy.

The second class of solutions corresponds to $\Delta=0(P^3=0)$
and is given by
$$
\eqalign{
&(A^{0})^{2}(\tau, \sigma)=((P^{1})^{2}+(P^{2})^{2})(\sigma)\tau ^{2}
+2P^{0}(\sigma)\tau +I^{0}(\sigma), \quad
A^{1,2}(\tau, \sigma)=P^{1,2}(\sigma)\tau +I^{1,2}(\sigma)\cr
&A^{3}(\tau, \sigma)=I^{3}(\sigma)\cr}
\numbereq\name{\eqznvbo}
$$
while the constraints read
$$
(P^{0})^{2}=I^{0}((P^{1})^{2}+(P^{2})^{2}), \quad
P^{1}(I^{1})^{'}+P^{2}(I^{2})^{'}={{P^0}{I^0}'\over {2I^0}}.
\numbereq\name{\eqxklo}
$$
The form of this solution suggests that string propagation
has been restricted to two space dimensions.
The energy-momentum tensor for the null string is obtained
by varying the action with respect to the metric $G_{\mu\nu}(X)$ at the
spacetime point $X$
$$
\sqrt{-G}T^{\mu\nu}(A)= {1\over {\lambda}}
\int d\tau d\sigma \dot{A}^{\mu}\dot{A}^{\nu}
\delta ^{(4)}(A-A(\tau ,\sigma ))
\numbereq\name{\eqaxnc}
$$
where $A^{\mu}(\tau, \sigma)$ are the string coordinates.
Furthermore we integrate $T^{\mu\nu}(A)$ over a spatial volume
thus enclosing the string at fixed time $A^{0}$ as it was proposed in
\ref{\fgh}{H. J. de Vega and N. Sanchez, \imp7 (1992) 3043.},
$$
I^{\mu \nu }(A^{0})=\int \sqrt{-G}T^{\mu \nu }(A)d^{3}{\vec A}.
\numbereq\name{\eqandgi}
$$
If we insert Eqn. (\eqohdp) into Eqn. (\eqandgi) we find
the following expressions for the energy and the momentum
of the null string
$$
I^{00}={1\over {\lambda}}\int d{\sigma} {\sqrt{{{(P^{0})^2-(P^{3})^2}\over
{I^{0}}}+{(P^{3})^2\over {(A^{0})^2}}}}, \quad
I^{0i}={1\over {\lambda}}\int d{\sigma}P^{i}(\sigma),
\quad I^{03}={1\over {\lambda}}\int d{\sigma}{{P^{3}(\sigma)}\over
{(A^{0})^2}}
\numbereq\name{\eqancoh}
$$
with $i=1,2$. Similarly we find for the components of the pressure
$$
I^{ii}={1\over {\lambda}}\int d{\sigma}{{(P^i)^2}\over
{\sqrt{{{(P^{0})^2-(P^{3})^2}\over
{I^{0}}}+{(P^{3})^2\over {(A^{0})^2}}}}}, \quad
I^{33}={1\over {\lambda}}\int d{\sigma}{{(P^3)^2}\over
{{(A^{0})^4}{\sqrt{{{(P^{0})^2-(P^{3})^2}\over
{I^{0}}}+{(P^{3})^2\over {(A^{0})^2}}}}}}
\numbereq\name{\eqadnvo}
$$
while the components $I^{i3} (i=1, 2)$ and $I^{12}$
$$
I^{i3}={1\over {\lambda}}\int d{\sigma}{{(P^iP^3)}\over
{A^{0}{\sqrt{{{(P^{0})^2-(P^{3})^2}{(A^{0})^2}\over
{I^{0}}}+{(P^{3})^2\over {(A^{0})^2}}}}}}, \quad
I^{12}={1\over {\lambda}}\int d{\sigma}{{(P^1P^2)}\over
{\sqrt{{{(P^{0})^2-(P^{3})^2}\over
{I^{0}}}+{(P^{3})^2\over {(A^{0})^2}}}}}
\numbereq\name{\eqabvvo}
$$
represent stresses which act on the string.
We observe that the isotropic pressure components $I^{ii} (i=1,2)$
attain constant values as $A^{0} \mapsto {\infty}$ while the anisotropic
component $I^{33}$ tends to zero.
If we now let a sequence of strings to move towards the anisotropy
with $\left({{dN}\over {dA^{0}}}\right)=N_{0}$
for a constant number of strings that move from the
origin $(A^{3}=0)$ towards the anisotropy we have
$$
{{dW}\over {dA^{0}}}=N_{0}{{dW}\over {dN}}=N_{0}I^{00}.
\numbereq\name{\eqwxiou}
$$
Integrating from the initial cosmic time $(A^{0}=\sqrt{I^{0}})$
to $(A^{0}=\bar{A}^{0})$ one obtains an expression for the
energy of $N$ strings
$$
\eqalign{
W&={{N_0}\over {2{\lambda}}}\int d{\sigma}
\lbrack\sqrt{(P^{3})^{2}+((P^{1})^2+(P^{2})^2)(A^{0})^{2}}\cr
&-P^{3}
ln\left({{P^{3}+\sqrt{(P^{3})^{2}+((P^{1})^2+(P^{2})^2)
(A^{0})^{2}}}\over {{\sqrt{((P^{1})^2+(P^{2})^2)}}
A^{0}}}\right)\rbrack_{\sqrt{I^{0}}}^{\bar{A}^{0}}\cr}
\numbereq\name{\eqacnpr}
$$
The distribution of strings thus gains energy that grows linearly with
respect to the cosmic time $(A^{0})$. On the contrary if we constrain
the strings to move only along the anisotropic direction the energy
grows as $W \sim {\ln A^{0}}$.
Therefore in the case
in which one would allow for the backreaction,
the strings would gain energy from the gravitational
field, thereby altering the anisotropy.

\newsection First Order Corrections.
Let's now discuss the first order corrections $B^{\mu}
(\tau, \sigma)$. The equations of motion
become
$$
\eqalign{
&\ddot{B}^{0}+2A^{0}(\dot{A}^{3}\dot{B}^{3})+(\dot{A}^{3})^{2}B^{0}=
(A^{0})^{''}+A^{0}(A^{'3})^{2}, \quad
\ddot{B}^{1,2}=(A^{1,2})^{''}\cr
&\ddot{B}^{3}+{2\over {A^{0}}}
(\dot{A}^{0}\dot{B}^{3}+\dot{A}^{3}\dot{B}^{0})-
{{2\dot{A}^{0}\dot{A}^{3}B^{0}}\over {(A^{0})^{2}}}=
(A^{3})^{''}+{2\over {A^{0}}}(A^{0})^{'}(A^{3})^{'}\cr}
\numbereq\name{\eqsbci}
$$
and the constraints
$$
\eqalign{
&-(\dot{B}^{0}A^{'0}+\dot{A}^{0}B^{'0})+(\dot{B}^{1}A^{'1}
+\dot{A}^{1}B^{'1})+(\dot{B}^{2}A^{'2}+\dot{A}^{2}B^{'2})\cr
&+(A^{0})^{2}(\dot{B}^{3}A^{'3}+\dot{A}^{3}B^{'3})
+2A^{0}\dot{A}^{3}A^{'3}B^{0}=0\cr
&-2\dot{A}^{0}\dot{B}^{0}+2\dot{A}^{1}\dot{B}^{1}
+2\dot{A}^{2}\dot{B}^{2}+2(A^{0})^{2}\dot{A}^{3}\dot{B}^{3}
+2A^{0}(\dot{A}^{3})^{2}B^{0}\cr
&-(A^{'0})^{2}+(A^{'1})^{2}+(A^{'2})^{2}+(A^{0})^{2}(A^{'3})^{2}=0.\cr}
\numbereq\name{\eqabvm}
$$
Two of these equations can be integrated directly giving
$$
B^{1,2}(\tau, \sigma)={1\over 6}(P^{1,2})^{''}(\sigma)\tau ^{3}
+{1\over 2}(I^{1,2})^{''}(\sigma)\tau ^{2}+
Q^{1,2}(\sigma)\tau +J^{1,2}(\sigma).
\numbereq\name{\eqnbxoi}
$$
By multiplying the equation of motion for $B^{3}(\tau, \sigma)$
with ${A^{0}}^2$ and taking into account Eqn. (\eqbmoc) we find
$$
{{\partial}\over {\partial \tau}}\left[(A^{0})^{2}\dot{B}^{3}+
2A^{0}B^{0}\dot{A}^{3}\right]=
{{\partial}\over {\partial \sigma}}\left[(A^{0})^{2}(A^{3})^{'}\right]
\numbereq\name{\eqahfl}
$$
As a result there exists a function $f=f(\tau ,\sigma )$ such that
$$
(A^{0})^{2}\dot{B}^{3}+2A^{0}B^{0}\dot{A}^{3}
={{\partial f}\over {\partial \sigma}}, \quad
(A^{0})^{2}(A^{3})^{'}={{\partial f}\over {\partial \tau }}
\numbereq\name{\eqvboxl}
$$
Using Eqn. (\eqohdp) we find
$$
\eqalign{
f(\tau ,\sigma)&=(I^{3})^{'}[{1\over 3}((P^1)^{2}+(P^2)^2)\tau ^{3}
+P^{0}\tau ^{2}+I^{0}\tau +J^{0}]\cr
&+{(P^{3})\over ((P^1)^{2}+(P^2)^2)}
\left[((P^1)^{2}+(P^2)^2)\tau ^{2}+P^{0}\tau +Q^{0}
\right]^{'}+f^{0}\cr}
\numbereq\name{\eqaxhvb}
$$
where $J^{0},Q^{0},f^{0}$ are arbitrary functions of $(\sigma )$.
Finally by combining Eqns. (\eqsbci), (\eqabvm) and (\eqvboxl) we
arrive at the following expression
$$
\eqalign{
&{{\partial}\over {\partial \tau}}\left[(A^{0})^{3}\dot{B}^{0}\right]
=-{3\over 2}(A^{0})^{2}[
(A^{'0})^{2}-(A^{'1})^{2}-(A^{'2})^{2}-(A^{0})^{2}(A^{'3})^{2}]\cr
&+3(A^{0})^{2}[\dot{A}^{1}\dot{B}^{1}+\dot{A}^{2}\dot{B}^{2}]+
P^{3}({{\partial f}\over {\partial\sigma}})
+(A^{0})^{2}[A^{0}(A^{0})^{''}+(A^{0})^{2}(A^{'3})^{2}]=h
(\tau, \sigma)\cr}
\numbereq\name{\eqdbvoj}
$$
Upon integration we derive an explicit expression for
$B^{0}(\tau, \sigma)$
$$
B^{0}(\tau, \sigma)={\int^{\tau}}d{\tau'}{1\over {{A^{0}
(\tau', \sigma)}^{3}}}{\int^{\tau'}}d{\tilde{\tau}}h({\tilde\tau}, \sigma)
+{\int^{\tau}}d{\tau'}{{g(\sigma)}\over {{A^{0}(\tau', \sigma)}
^{3}}}+l(\sigma)
\numbereq\name{\eqbcovj}
$$
Finally by substituting $B^{0}$ into Eqn. (\eqvboxl)
and integrating we find an expression for
$B^{3}(\tau, \sigma)$
$$
B^{3}(\tau, \sigma)={\int^{\tau}}d{\tilde{\tau}}{{f'({\tilde{\tau}},
{\sigma})}\over {{A^{0}(
{\tilde{\tau}}, \sigma)}^{2}}}-2{\int^{\tau}}d{\tilde{\tau}}{{P^3({\sigma})}
\over {{A^{0}(
{\tilde{\tau}}, \sigma)}}}+m(\sigma)
\numbereq\name{\eqbvnoogj}
$$
The general solution $B^{\mu}(\tau, \sigma)$
is complicated and
depends on arbitrary functions of $\sigma$ which determine the
initial shape and momentum of the string.
In order to simplify the calculation we will consider some special cases.
Initially we will assume that the
string at $\tau=0$ forms a circle of radius $R$ in the
$X^1-X^3$ plane and furthermore that all the points of the
string move with the same momentum $P^1$ parallel to the $X^1$
axis, more specifically
$$
I^{1}(\sigma)=R{\sin{\sigma}}, \quad I^{3}(\sigma)=R{\cos{\sigma}},
\quad P^{1}\ne 0, \quad P^{2}=P^{3}=0.
\numbereq\name{\eqasiuov}
$$
From the constraints Eqn. (\eqxbvo) we find that $(P^{0})^2=
I^{0}(P^{1})^2$ and $I^{0}=R^2{\sin^{2}\sigma}$. The overall
solution for the $A's$ becomes
$$
A^{0}(\tau, \sigma)=(P^{1}{\tau}+R{\sin{\sigma}}), \quad
A^{1}(\tau, \sigma)=P^{1}{\tau}+R{\sin{\sigma}}, \quad
A^{2}(\tau, \sigma)=0, \quad A^{3}(\tau, \sigma)=R{\cos{\sigma}}.
\numbereq\name{\eqaohfk}
$$
It is also interesting to calculate the distance of the string
between $(\sigma, \tau)$ and $(\sigma+d\sigma, \tau)$. We find
that for the solution (\eqaohfk)
$$
ds^2=(P^{1}R{\tau}{\sin{\sigma}}+R^2{\sin^{2}{\sigma}})^2d{\sigma^2}
=(A^{0})^2R^2{\sin^{2}{\sigma}}d{\sigma^2}
\numbereq\name{\eqanbvi}
$$
We observe that the string is expanding at the same rate as
the spacetime.
The solution then to the Eqn. (\eqsbci), using Eqn. (\eqaohfk) is
$$
\eqalign{
&B^{0}(\tau, \sigma)=-R{\sin{\sigma}}{{\tau^2}\over 2}+P^{1}R^2
{\sin^{2}{\sigma}}{{\tau^3}\over 6}+R^3{\sin^{3}{\sigma}}
{{\tau^2}\over 2}+H^{0}(\sigma){\tau}+Q^{0}(\sigma)\cr
&B^{1}(\tau, \sigma)=-R{\sin{\sigma}}{{\tau^2}\over 2}+
H^{1}(\sigma){\tau}+Q^{1}(\sigma) \quad B^{2}(\tau, \sigma)=
H^{2}(\sigma){\tau}+Q^{2}(\sigma)\cr
&B^{3}(\tau, \sigma)=-R{\cos{\sigma}}{{\tau^2}\over 6}
-{{4R^2}\over {3P^1}}({\cos{\sigma}}{\sin{\sigma}}){\tau}
-\lbrack {{4R^4{\cos{\sigma}}{\sin^{3}{\sigma}}} \over
{3(P^1)^2}}+{{H^{3}(\sigma)}\over {(P^1)}} \rbrack
{(P^1{\tau}+R{\sin\sigma})^{-1}}\cr
&+Q^{3}(\sigma)\cr}
\numbereq\name{\eqaxnv}
$$
while the constraints Eqn. (\eqabvm) imply that the functions
$H^{\mu}(\sigma)$ and $Q^{\mu}(\sigma)$ satisfy the following relations
$$
2P^{1}(H^{0}-H^{1})=R^{4}{\sin^{4}{\sigma}}, \quad
P^{1}(Q^{0'}-Q^{1'})+R{\cos{\sigma}}(H^{0}-H^{1})+R{\sin{\sigma}}=0
\numbereq\name{\eqapnzi}
$$
Similarly we can assume that the
string at $\tau=0$ forms a circle of radius $R$ in the
$X^2-X^3$ plane and furthermore that all the points of the
string move with the same momentum $P^3$ parallel to the $X^3$
axis, the anisotropy direction, more specifically
$$
I^{1}=0, \quad
I^{2}(\sigma)=R{\cos{\sigma}}, \quad
I^{3}(\sigma)=R{\sin{\sigma}}, \quad P^{3}\ne 0, \quad P^{1}=P^{2}=0.
\numbereq\name{\eqabzo}
$$
The constraints then imply that $(P^{0})^2=(P^{3})^2$ and $I^{0}(\sigma)
=e^{2R{\sin{\sigma}}}$ while the overall solution reads
$$
\eqalign{
&(A^{0})^{2}(\tau, \sigma)=2{P^{3}}\tau+e^{2R{\sin{\sigma}}},
\quad A^{1}(\tau, \sigma)=0,
\quad A^{2}(\tau, \sigma)=R{\cos{\sigma}} \cr
&A^{3}(\tau, \sigma)={1\over 2}
{\ln{(2P^{3}{\tau}+e^{2R{\sin{\sigma}}})}}\cr} 
\numbereq\name{\eqzcxpo}
$$
For this particular solution we find that the distance of the string
between $(\sigma, \tau)$ and $(\sigma+d\sigma, \tau)$ is
$$
ds^2=R^2{\sin^{2}{\sigma}}d{\sigma^2}
\numbereq\name{\eqbjhvi}
$$
which implies that the cosmological expansion balances exactly
the string contraction.
The first order corrections $B^{0}(\tau, \sigma)$ and $B^{3}(\tau, \sigma)$
become
$$
\eqalign{
&B^{0}(\tau, \sigma)=C^{0}(\sigma)(2{P^{3}}\tau
+e^{2R{\sin{\sigma}}})^{3\over 2}+D^{0}(\sigma)
(2{P^{3}}\tau
+e^{2R{\sin{\sigma}}})^{1\over 2}\cr
&+H^{0}(\sigma)
(2{P^{3}}\tau
+e^{2R{\sin{\sigma}}})^{-{1\over 2}}+Q^{0}(\sigma)\cr
&B^{3}(\tau, \sigma)=C^{3}(\sigma)(2{P^{3}}\tau
+e^{2R{\sin{\sigma}}})+D^{3}(\sigma){\ln{
(2{P^{3}}\tau
+e^{2R{\sin{\sigma}}})}}\cr
&+H^{3}(\sigma)
(2{P^{3}}\tau
+e^{2R{\sin{\sigma}}})^{-{1\over 2}}+Q^{3}(\sigma)\cr}
\numbereq\name{\eqafnbc}
$$
where $C(\sigma)$ and $D(\sigma)$ are known functions
of $e^{2R{\sin{\sigma}}}$ and $P^3$ and $H(\sigma), Q(\sigma)$
are arbitrary functions of $\sigma$. The other components
$B^{1, 2}(\tau, \sigma)$ retain their previous form.
The form of the solution $X^{\mu}=A^{\mu}+c^2B^{\mu}$ suggests that in
the null string expansion the overall motion of the string is treated exactly
while the internal motion is approximated by a $\tau$-expansion.

Finally we would like to comment on the algebra of the constraints
and the quantization of the null
string. The constraints
$$
G(\sigma)=(P^{0})^{2}-I^{0}((P^{1})^{2}+(P^{2})^{2})-(P^{3})^{2}, \quad
F(\sigma)=P^{1}(I^{1})^{'}+P^{2}(I^{2})^{'}+P^{3}(I^{3})^{'}
-{{P^0}{I^0}'\over {2I^0}}
\numbereq\name{\eqaghnc}
$$
generate reparametrizations of the
form $\tau=f({\tilde{\tau}}, {\tilde{\sigma}})$,
$\sigma=h({\tilde{\sigma}})$ and obey the following Poisson brackets
$$
\eqalign{
\{ F(\sigma), F(\sigma') \}&=
2F(\sigma'){\dpr}-F'(\sigma'){\d}, \qquad \{ G(\sigma), G(\sigma') \}=0,\cr
\{ G(\sigma), F(\sigma') \}&=2G(\sigma'){\dpr}-G'(\sigma'){\d}.\cr}
\numbereq\name{\eqacvoi}
$$
This is the classical algebra for the constraints of the null
string
\ref{\spl}{R. Amorim and J. Barcelos-Neto, \pl253 (1991) 313.}
and it arises as the Inonu-Wigner contraction $( T \mapsto 0 )$
\ref{\inw}{E. Inonu and E. Wigner, {\it Proc. Nat. Acad. Sci. (US)}
{\bf 39} (1953) 510.} of the algebra of the constraints of the tensile string.
Quantum mechanically the Poisson brackets will be replaced
by commutators and this offers the possibility for the emergence
of central extension in the algebra
\ref{\lizzi}{F. Lizzi, B. Rai, G. Sparano and A. Srivastava,
\pl182 (1986) 326, F. Lizzi, \mpl9 (1994) 1465, J. Gamboa, C. Ramirez
and M. Ruiz-Altaba, \np338 (1990) 143, J. Isberg, U. Lindstrom,
B. Sundborg and G. Theodoridis, \np411 (1994) 122, A. Nicolaidis,
J. Paschalis and P. Porfyriadis, \pr58 (1998) 047901.}.
We hope to report our results towards this direction in a
future communication.

\newsection Conclusions.
In this section we will briefly recapitulate what we have done
in this paper. We studied string propagation in an anisotropic
Milne background. The equations of motion and the constraints
were solved by means of the null string expansion. In this
scheme the string equations of motion and the constraints are
systematically expanded in powers of the world-sheet speed of
light $c^2$ which is proportional to the string tension $T$.
To zeroth order the string is described by a null (tensionless)
string-an extended object. The points of the null string interact
only with the gravitational background. The first order correction
introduces the string tension $T$ and thus the
self-interaction among the points of the string. We derived exact analytical
expressions for the zeroth and first order solutions. We also
calculated the stress tensor for the null string and examined
some of its cosmological implications.

{\bf Acknowledgments.} \vskip .01in \noindent
We would like to thank J. Liu for useful discussions.
I. G. was supported in part by the Department of Energy under Contract
Number DE-FG02-91ER40651-TASK B and by NSF
under grant PHY-9722083.
K. K. and A. K. would like to thank the Greek State Scholarships
Foundation for the financial support during this work.
This work was also supported by the PENED GGET No 672/A.U.th (1768)
``Nonlinear
Theories of Gravitation in Modern Astrophysics and Microcosmology''.

\immediate\closeout1
\bigbreak\bigskip

\line{\twelvebf References. \hfil}
\nobreak\medskip\vskip\parskip

\input refs

\vfil\end

\bye